\documentclass[numbered]{trbunofficial}
\usepackage{graphicx}

\usepackage{hyperref}
\hypersetup{hidelinks}
\usepackage{rotfloat}
\floatstyle{ruled}
\newfloat{algorithm}{tbp}{loa}
\floatname{algorithm}{Algorithm}
\usepackage{subcaption}
\usepackage{tikz}
\usetikzlibrary{arrows, decorations.markings}
\tikzstyle{vertex}=[circle, draw, inner sep=0pt, minimum size=6pt]
\newcommand{\vertex}{\node[vertex]}

\AuthorHeaders{Jing Lu and Carolina Osorio}
\title{On the approximation of queue-length distributions in transportation networks}

\author{%
  \textbf{Jing Lu}\\
  Operations Research Center\\
  Massachusetts Institute of Technology, USA\\
  \hfill\break
  \textbf{Carolina Osorio}\\
  Department of Decision Sciences\\
  HEC Montr\'{e}al, Canada\\
  \hfill\break%
  \textbf{Jing Lu, Ph.D., Corresponding Author}\\
  jl3724@mit.edu
}

\begin{document}
\maketitle

\section{Abstract}

This paper focuses on the analytical probabilistic modeling of vehicular traffic. It formulates a stochastic node model. It then formulates a network model by coupling the node model with the link model of \citep{OsoJing17}, which is a stochastic formulation of the traffic-theoretic link transmission model. The proposed network model is scalable and computationally efficient, making it suitable for urban network optimization. For a network with $r$ links, each of space capacity  $\ell$, the model has a complexity  of $\mathcal{O}(r\ell)$. The network model yields the marginal distribution of link states. 
The model is validated versus a simulation-based network implementation of the stochastic link transmission model. The validation experiments consider a set of small network with intricate traffic dynamics. For all scenarios, the proposed model accurately captures the traffic dynamics. 
The network model is used to address a signal control problem. Compared to the probabilistic link model of \citep{OsoJing17} with an exogenous node model and a benchmark deterministic network loading model, the proposed network model derives signal plans with better performance.
The case study highlights the added value of using between-link (i.e., across-node) interaction information for traffic management and accounting for stochasticity in the network.

\hfill\break%
\noindent\textit{Keywords}: Probabilistic node modeling, Computational efficiency, Stochastic network loading, Network optimization
\newpage

\section{Introduction}
This paper formulates an analytical stochastic (i.e., probabilistic) network model. It formulates a node model, and couples the node model with an existing stochastic link model. Our focus on stochastic analytical modeling is motivated by an interest in developing macroscopic traffic models that can both provide a more detailed distributional description of network performance and can be used for large-scale network analysis and optimization.
The increase in the quantity, quality and resolution of traffic data available allows to validate models that provide a probabilistic description of traffic. 
Such models can be used to enhance the reliability and the robustness of our transportation networks. Major agencies in the US and in Europe are focusing on measuring and improving such reliability and robustness performance metrics. 
Our research in probabilistic network modeling is to formulate traffic models that 
are suitable for the efficient analysis and optimization of large-scale networks.

A stochastic network model consists of two main components: a link model, which describes traffic dynamics within a homogeneous road segment, and a node model, which describes traffic dynamics between links.  There is an increasing and recent interest in the formulation of probabilistic traffic-theoretic link models.  Recent reviews include \citep{JabariPhD12} and \citep{Laval14}.  Probabilistic link models formulated with traffic-theoretic foundations include  \cite{Boel06,Sumalee11,Jabari12,Jabari13,OsoFloBie11,Deng13,Laval14,Laval15,OsoFlo15_TS,OsoJing17,lu2020probabilistic}. The main approaches have been the introduction of randomness in the cell transmission model, the extensions of the variational theory of \citep{Daganzo05} and the use of probabilistic queueing theory \cite{olszewski1994modeling,Heidemann01,viti2010probabilistic,OsoFloBie11,OsoFlo15_TS,OsoJing17}. Another stream of research in the estimation of queue length distribution is based on observed data collected from probe vehicles \cite{comert2009queue, comert2011analytical, comert2013simple, comert2016queue}.

The model of \citep{OsoFlo15_TS} is a queueing-theoretic model that is a stochastic formulation of the deterministic link-transmission model of \citep{Yperman07}, which itself is an operational formulation of Newell's simplified theory of kinematic waves \cite{Newell93}. For a link with space capacity $\ell$, the model complexity is in the order of $\mathcal{O}(\ell^3)$. \citep{OsoJing17} have extended the formulation of \citep{OsoFlo15_TS}, leading to a model with complexity in the order of $\mathcal{O}(\ell)$. This is a formulation that scales linearly with the link's space capacity, making it efficient and appropriate for large-scale network analysis. The model in Chapter 3 of \citep{lu2020probabilistic} further reduces the model complexity to a constant that no longer depends on the link’s space capacity. This makes the proposed model suitable for large-scale network optimization or situations where a large number of model evaluations is required. In this paper, we focus on the network model development with the analytical probabilistic link model of \citep{OsoJing17} where the transient distribution of the queue-length is produced.

The formulation of traffic-theoretic node models is limited to deterministic formulations
(e.g., \cite{daganzo1995finite,lebacque1996godunov,lebacque2005first,corthout2012non,flotterod2011operational,tampere2011generic,smits2015family}). These node models satisfy supply and demand restrictions at the link boundaries, mass balance requirements on flows across the node and some further considerations in the distribution of demands to different outgoing links to reflect destination desires of drivers. Most deterministic node models impose flow maximization to facilitate the physical requirement that traffic-holding does not exist. It is shown in a recent work of \citep{jabari2016node} that flow maximized solutions are only sufficient conditions for holding-free solutions, especially in complex congested urban intersections. Other themes of node modeling include the derivation of formulas for calculating node flows that satisfies the aforementioned restrictions \cite{bliemer2007dynamic}. Nevertheless, one challenge of network modeling that persists today is the analytically intractability due to the coupling with link dynamics and node models that are typically stated as optimization problems. 

The development of stochastic node model is quite limited and mostly focused on simple network topology.
The model of \citep{OsoFloBie11} includes a stochastic node model for a simple two queue network. \citep{OsorioYamani_forthcoming} proposes an analytical model to approximate the transient aggregate joint queue-length distribution of tandem links, the aggregation-disaggregation technique and network decomposition methods are used to reduce the model complexity. \citep{OsorioWang_submitted} propose an analytical approximation of the aggregate stationary joint queue-length distributions for arbitrary size and it can be extended to general topology by coupling with the method introduced in \citep{OsorioEJOR}.\citep{FloOso_DTA} develops an analytical stochastic link transmission model (SLTM) for networks that produces the transient joint queue-length distribution and demonstrated its usage on linear, diverge and merge node models within the SLTM framework. Nevertheless, there is a lack of stochastic node models that apply for general network topology and can be integrated with SLTM for potentially large-scale network analysis.

In this article, we formulate a general stochastic node model that can be combined with various types of stochastic link models for efficient network analysis. We couple the proposed stochastic node model with the link model of \citep{OsoJing17}.
Section \ref{model} formulates the proposed stochastic network model. The model is validated in Section \ref{validation}. It is used to address a signal control problem in Section \ref{casestudy}. This case study shows the added value of accounting for between-link (i.e., across node) interactions for traffic management.
Conclusions and future work are presented in Section \ref{conclusion}.

\section{Network model formulation}\label{model}
We present a general node model formulation. The proposed node model can be coupled with a variety of probabilistic link models to yield a network model.  
We then couple the node model with the link model of \citep{OsoJing17}, which produces the transient distribution of the link's queue length distribution. Section~\ref{link_model} briefly describes the link model, which is, hereafter, referred to as  the mixture model. The node model  formulation is given in Section~\ref{node_model}.

\subsection{Link model}\label{link_model}
We briefly outline here the main ideas of the link model of \cite{OsoJing17}. For a more detailed description, we refer the reader to \cite{OsoJing17}. The mixture model considers an isolated single-lane link. It assumes a triangular fundamental diagram. The link is parameterized as follows: free flow velocity $v$, backward wave speed $w$ (negative), flow capacity $\hat{q}$, jam density $\hat{\varrho}$, and link length $L$.
The link model approximates the following process. Upon arrival to the link, vehicles are delayed 
$L/v$ time units (known as the forward lag). After incurring this delay, vehicles are then ready for departure at the downstream end of the link.  
Upon departure from the link, there is a delay of $L/|w|$ time units (known as the backward lag) before the newly available space is made available at the upstream end of the link. These delays or lags capture the time it takes forward, and backward, kinematic waves to traverse the link.
The mixture model is a continuous-space discrete-time model. The lags are rounded to the nearest integer: $L/(v\delta)$ is rounded to $k^{\text{fwd}}$, and $L/(|w|\delta)$  is rounded to $k^{\text{bwd}}$, where $\delta$ is the incremental time.

The vehicular flow process described above is modeled with two queues.
Vehicles that are ready for departure enter the \textit{downstream queue}, denoted $DQ$. In other words, $DQ$ captures the downstream boundary conditions of the link. Spaces that are occupied (i.e., they are either occupied by vehicles on the link, or by vehicles that have departed the link less than $L/(|w|\delta)$ time units ago) are represented by the \textit{upstream queue}, denoted $UQ$. In other words, $UQ$ captures the upstream boundary conditions of the link.

The model of \cite{OsoJing17} considers two univariate independent models: a univariate $UQ$ model, which captures the link's upstream boundary conditions, and a univariate $DQ$ model, which captures the link's downstream boundary conditions. Each of these two models approximates two distributions: (i) the distribution of the link's upstream queue $UQ$, and (ii) the distribution of the link's downstream queue $DQ$.  

Then, the model of \cite{OsoJing17} is defined as a mixture of these two models.
For this discrete time model, let $UQ(k)$ denote the number of vehicles (or spaces occupied) in $UQ$ at the end of discrete time interval $k$. We use similar notation for the downstream queue $DQ(k)$.
The mixture model yields the marginal distributions of the downstream queue and of the upstream queue. These are given by:
\begin{equation}
P(UQ(k)) = \hat{w} P^{UQ}(UQ(k))+ (1-\hat{w}) P^{DQ}(UQ(k)) 
\label{eq:mix1}
\end{equation}
\begin{equation}
P(DQ(k)) = \hat{w} P^{UQ}(DQ(k))+ (1-\hat{w}) P^{DQ}(DQ(k)), 
\label{eq:mix2}
\end{equation}
where $P^{UQ}(UQ(k))$ (resp. $P^{DQ}(UQ(k))$) denotes the $UQ$ distribution obtained from the univariate $UQ$ (resp. $DQ$) model, 
$P^{UQ}(DQ(k))$ (resp. $P^{DQ}(DQ(k))$) denotes the $DQ$ distribution obtained from the univariate $UQ$ (resp. $DQ$) model,
and $\hat{w}$ denotes an exogenous weight parameter calculated according to Equation (37) of \cite{OsoJing17}. 

The mixture model has linear complexity in the links space capacity $\ell$ because the two univariate models are independent. In other words, the distributions $P^{UQ}(UQ)$ and $P^{DQ}(UQ)$ are evaluated independently. Similarly, the distributions $P^{UQ}(DQ)$ and $P^{DQ}(DQ)$ are evaluated independently. 

Each univariate model is then formulated as a system of nonlinear equations. For a given time interval $k$, the expected inflow into, and the expected outflow from, the link are assumed constant, and the solution of the system of nonlinear equations yields the transient (i.e., time-dependent) distribution of $UQ$ (or of $DQ$) within the time interval $k$ of duration $\delta$. At the end of each time interval, the expected inflows and outflows are then updated and used for the next time interval.

The stochastic processes that govern the link dynamics are as follows. The link is assumed to have an inhomogeneous Poisson arrival process with exogenous arrival rate $\lambda(k)$, and independent and exponentially distributed service times at the downstream end of the link with exogenous downstream flow capacity $\mu(k)$. It is shown in Chapter 3 section 3.4.2 of \cite{lu2020probabilistic} that with properly calibrated time-varying arrival rate and service rate (i.e., $\lambda(k)$ and $\mu(k)$) and time increment $\delta$, a link's boundary conditions for realistic traffic situation such as signalized links and platoon arrival patterns can be estimated accurately compared to a details microscopic traffic simulator.

\subsection{Node model}\label{node_model}
We now formulate the node model. The analysis hereafter is for a given discrete time interval $k$. Hence, we use interchangeably $UQ$ and  $UQ(k)$, as well as $DQ$ and $DQ(k)$. 
Consider a general node with $m$ upstream links and $n$ downstream links, let $M$ be the set of upstream links and $N$ be the set of downstream links. We assume exogenous turning probabilities $\{p_{ij}, i\in M, j \in N\}$. Each link $i$ has an exogenous downstream flow capacity $\mu_i$ and space capacity $\ell_i$. The rate at which vehicles can start trips at link $i$ is exogenous and is denoted $\gamma_i$.  

For a given upstream link $i$ ($i \in M$), the number of vehicles ready for departure is $DQ_i$. For a given downstream link $j$ ($j \in N$), the number of spaces available at the upstream end of link $j$ (to accommodate demand from upstream links) is $(\ell_j-UQ_j)$. 

We assume that a vehicle from $i$ can transition to $j$ if the following two conditions hold. First, the link $j$ is not full so that there is space available to receive demand from upstream, i.e., $UQ_j< \ell_j$. Second, the vehicle is not blocked by vehicles in front that intend to leave link $i$ to other downstream links other than link $j$. A sufficient, but not necessary, condition to ensure that vehicles from link $i$ can transition to link $j$ is that there is space available to receive vehicles in all downstream links, i.e., $\forall n \in N,  UQ_n< \ell_n$. Additionally, if downstream links are not full, then the \textbf{instantaneous} departure rate from link $i$ corresponds to its flow capacity, $\mu_i$. This leads to the following expression for the  expected flow rate from link $i$ to link $j$ during time interval $k$ of length $\delta$:
\begin{equation}
q_{i\rightarrow j}^{out}(k)=p_{ij}\mu_i P(DQ_i(k)>0,UQ_n(k)< \ell_n \ \forall n \in N).
\label{eq:node}
\end{equation} 

This proposed approximation may underestimate the true expected flow rate, since $\{ DQ_i(k)>0$, $UQ_n(k)< \ell_n \ \forall n \in N\}$ is a sufficient, but not a necessary, condition for flow transmission. For instance, there are cases where a downstream link is full, yet it does not hinder outflow from all upstream links. This can happen if there are no vehicles upstream that intend to enter the full downstream link. 
However, since the it is assumed that there is always a constant non-zero proportion of outflows $p_{ij}$ from upstream link $i$ intent to enter the downstream link $j$ that is full, link $i$ will eventually be blocked, e.g., eventually the most downstreamed vehicle will become one that intends to join link $j$ but not able to and hence blocked the outflow of the single-lane link $i$.
The validation results of the next section indicate, this approximation has a  minor effect on the accuracy of the model.

Combining the above expression with a link model requires the computation of the joint probability $P(DQ_i(k)>0,UQ_n(k)< \ell_n \ \forall n \in N)$.
In general, a straightforward, but also inefficient, way to compute this is to formulate a network model that jointly tracks the state of all links in the network. For a network with $r$ links, each with space capacity $\ell$, the model complexity would be in the order of $\mathcal{O}(\ell^r)$.
Instead, we propose an approach with complexity $\mathcal{O}(r\ell)$.
The main idea is to track the distribution of each link in the network independently, and to capture the dependencies between links by adjusting the arrival rates and the effective service rates of the links according to the expected flows within the network.

Section~\ref{network2} formulates an approximation for the above joint probability, $P(DQ_i>0,UQ_n< \ell_n \ \forall n \in N)$, which we refer to, hereafter, as the flow transmission probability. Section~\ref{network1} then  presents how the link model parameters (arrival rates and effective service rates) are approximated at every time step. Section~\ref{algorithm} summarizes the method, and presents it as an algorithm.

\subsubsection{Flow transmission probability}\label{network2}
In this section, we present an analytical approximation of the flow transmission probability that only requires knowledge of the marginal univariate distributions of $DQ$ and of $UQ$. 

We use the inclusive-exclusive principle (Section 2.1 of \cite{bjorklund2009set}) to rewrite $P(DQ_i>0,UQ_n< \ell_n \ \forall n \in N)$ as follows:
\begin{align}
P(DQ_i&>0,UQ_n< \ell_n \ \forall n \in N)= \label{eq:inclusive_exclusive}\\ \nonumber
1&-P(DQ_i=0)-\sum_{n \in N}P(UQ_n= \ell_n)\\ \nonumber
&+ \sum_{n \in N}P(DQ_i=0,UQ_n= \ell_n)+\sum_{n\neq m; n,m\in N}P(UQ_n= \ell_n,UQ_m=\ell_m)\\ \nonumber
&+...+(-1)^{|N|+1}P(DQ_i=0,UQ_n= \ell_n \ \forall n \in N). \nonumber
\end{align}
In Equation~\eqref{eq:inclusive_exclusive}, the distribution of a single link state (e.g., $P(DQ_i=0)$, $P(UQ_n= \ell_n)$) can be obtained directly from the link model. Let us now describe how we approximate the joint probabilities that appear in~\eqref{eq:inclusive_exclusive}.
Each joint probability is approximated by rewriting it as a conditional probability conditioned on the sum of the corresponding variables. For example:
\begin{align}
P(DQ_i&=0,UQ_n= \ell_n \ \forall n \in N)\label{eq:example}\\ \nonumber
&=P\left( DQ_i=0,UQ_n= \ell_n \ \forall n \in N \mid DQ_i+\sum_{\forall n \in N}UQ_n=\sum_{\forall n \in N}\ell_n\right) \\ \nonumber
&\cdot P\left(DQ_i+\sum_{\forall n \in N}UQ_n=\sum_{\forall n \in N}\ell_n\right).\nonumber
\end{align}
In order to derive an  approximate expression for the right-hand side of Equation~\eqref{eq:example}, we make two approximations.
First, we approximate $DQ_i+\sum_{\forall n \in N}UQ_n$ as a Poisson arrival process with  rate $q(k)$. 
Hence, 
\begin{equation}
P\left(DQ_i(k)+\sum_{\forall n \in N}UQ_n(k)=\sum_{\forall n \in N}\ell_n\right) = [\delta \cdot q(k)]^{\sum_{\forall n \in N}\ell_n} \frac{e^{-\delta \cdot q(k)}}{(\sum_{\forall n \in N}\ell_n)!},
\end{equation}
where $\delta$ is the time step length. The rate is given by the sum:
\begin{equation}
q(k)=  q^{DQ_i}(k)+\sum_{\forall n \in N}q^{UQ_n}(k),
\end{equation}
where $q^{DQ_i}(k)\delta$ (resp. $q^{UQ_n}(k)\delta$) represents the expected state of $DQ_i$ (resp. $UQ_n$) for discrete time interval $k$. These can be obtained directly from the link model (see Equations (7) and (12) of \cite{OsoJing17}).
Let us briefly explain how they are calculated. The expression for $q^{UQ_n}(k)$ is:
\begin{equation}
q^{UQ_n}(k) = \sum_{r=0}^{k-1} q_n^{\text{in}}(r) - \sum_{r=0}^{k-k^\text{bwd}-1} q_n^{\text{out}}(r),
\label{eq:qUQ}
\end{equation}
where $q_n^{\text{in}}(k)$ (resp. $q_n^{\text{out}}(k)$) denotes the instantaneous  inflow (resp. outflow) rate of link $n$ at the end of time interval $k$.  This expression is obtained from the observation that the expected state of $UQ_n$ represents the difference between the accumulated inflows to link $n$ during the last $k-1$ time steps and the accumulated outflows from link $n$ during the last $k-k^\text{bwd}-1$ time steps.
Similarly, the expression for $q^{DQ_i}(k)$ is:
\begin{equation}
q^{DQ_i}(k) = \sum_{r=0}^{k-k^\text{fwd} - 1} q_i^{\text{in}}(r) - \sum_{r=0}^{k-1} q_i^{\text{out}}(r).
\label{eq:qDQ}
\end{equation}
Equation~\eqref{eq:qDQ} considers the expected flow in $DQ_i$ as the difference between: (i) the sum of all of the expected inflows into  link $i$ from time $0$ to time $k-k^\text{fwd}-1$ (i.e., omitting the flows that have  still not arrived to the downstream end of the link)  and (ii) the sum of all expected outflows out of link $i$ (i.e., outflow from time  $0$  to time $k-1$).

Second, we approximate $DQ_i+\sum_{\forall n \in N}UQ_n$ as a sum of independent Poisson processes. Hence, the conditional variate $\{DQ_i,UQ_n, \ \forall n \in N\}$ given $\{DQ_i+\sum_{\forall n \in N}UQ_n=\sum_{\forall n \in N}\ell_n\}$ follows a multinomial distribution with parameters $\left(\sum_{\forall n \in N}\ell_n; q^{DQ_i}/q, q^{UQ_n}/q \ \forall n \in N\right)$ (cf., for instance, Section 2.12.4 of \cite{Larson81}). This leads to:
\begin{equation}
P\left( DQ_i =0,UQ_n= \ell_n \ \forall n \in N \mid DQ_i+\sum_{\forall n \in N}UQ_n=\sum_{\forall n \in N}\ell_n\right)
= \frac{(\sum_{\forall n \in N}\ell_n)!}{\prod_{n \in N}\ell_n !}\prod_{n \in N}\left(\frac{q^{UQ_n}}{q}\right)^{\ell_n}.
\end{equation}
Note that although we ignore the temporal dependencies among the queues, the multinomial distribution parameters ($q^{DQ_i}/q$ and $q^{UQ_n}/q$) are highly temporally dependent. 
With this same technique, we approximate each of the joint probabilities that appear in Equation \eqref{eq:inclusive_exclusive}.

\subsubsection{Link model arrival and effective service rates}\label{network1}
We now present how we use the expected flow  between links (Equation~\eqref{eq:node}) to approximate the arrival rates and the effective service rates of a link.

Consider a node with an upstream link $i$. The expected outflow rate from link $i$ is given by:
\begin{align}
q_i^{out}(k)&=\mu_i\left(1-\sum_{j\in N}p_{ij}\right)P(DQ_i(k)>0)+\sum_{j\in N}q_{i\rightarrow j}^{out}(k)\label{eq:qOut_def1}\\
&=\mu_i\left(1-\sum_{j\in N}p_{ij}\right)P(DQ_i(k)>0)+ \left(\sum_{j\in N}p_{ij}\right)\mu_iP(DQ_i(k)>0,UQ_n(k)< \ell_n \ \forall n \in N).\label{eq:qOut_def1b}
\end{align}
Equation \eqref{eq:qOut_def1} states that the expected outflow rate from link $i$ is the summation of the expected flow rate that leaves the network through  link $i$ and the  expected outflow from link $i$ to all its downstream links $j$. Equation~\eqref{eq:qOut_def1b} is obtained by inserting the expression for $q_{i\rightarrow j}^{out}(k)$ of Equation~\eqref{eq:node}.

Additionally, let $\hat{\mu}_i(k)$ denote the effective service rate of link $i$ at time step $k$. The effective service rate, $\hat{\mu}_i$, differs from the service rate, $\mu_i$, in that the former accounts for the impact of downstream spillbacks on the departure rate of link $i$, while the latter is an exogenous parameter that assumes the link is isolated and  represents its downstream flow capacity.

By definition, the expected outflow and the effective service rate are related as follows:
\begin{equation}
q_i^{out}(k) = \hat{\mu}_i(k) P(DQ_i(k) > 0)\label{eq:qOut_def2}.
\end{equation}
We can combine Equations~\eqref{eq:qOut_def1b} and \eqref{eq:qOut_def2} to obtain an expression for the effective service rate:
\begin{equation}
\hat{\mu}_i(k) = \frac{q_i^{out}(k)}{ P(DQ_i(k) > 0)}=\mu_i\left[\left(1-\sum_{j\in N}p_{ij}\right) + \left(\sum_{j\in N}p_{ij}\right)P(UQ_n(k)< \ell_n \ \forall n \in N \mid DQ_i(k)>0)\right].
\label{eq:effservice}
\end{equation}

Let us now describe how the expected flow between links is used to approximate the arrival rate of a link.
Consider a node with downstream link $j$. The expected inflow rate to link $j$ is given by: 
\begin{align}
q_j^{in}(k)&=\gamma_j P(UQ_j(k) < \ell_j) + \sum_{i\in M}q_{i\rightarrow j}^{out}(k)\label{eq:qIn_def1}\\
&=\gamma_j P(UQ_j(k) < \ell_j)+ \sum_{i\in M}p_{ij}\mu_iP(DQ_i(k)>0,UQ_n(k)< \ell_n \ \forall n \in N)\label{eq:qIn_def1b}
\end{align}
where $\gamma_j$ is the exogenous arrival rate to link $j$ (i.e., it represents the rate with which trips start at link $j$),
Equation \eqref{eq:qIn_def1}  states that the expected inflow to link $j$ is the summation of the expected flow entering the network through link $j$ and the expected flow from all upstream links $i$ to link $j$.

By definition, the expected inflow rate and the arrival rate, $\lambda_j(k)$, are related as follows:
\begin{equation}
q_j^{in}(k) = \lambda_j(k) P(UQ_j(k) < \ell_j).\label{eq:qIn_def2}
\end{equation}

We combine Equations~\eqref{eq:qIn_def1b} and \eqref{eq:qIn_def2} to obtain an expression for the arrival rate of link $i$ at time step $k$:
\begin{equation}
\lambda_j(k)=\frac{q_j^{in}(k)}{P(UQ_j(k) < \ell_j)}= \gamma_i + \sum_{i\in M}p_{ij}\mu_iP(DQ_i(k)>0,UQ_n(k)< \ell_n \ \forall n\neq j, n \in N \mid UQ_j(k) < \ell_j)
\label{eq:lambda}
\end{equation}

\subsubsection{Algorithm}\label{algorithm}
Algorithm \ref{algo1} summarizes the numerical evaluation of the proposed network model for all links $i$ in the set of  links $\mathcal{L}$.  The state distribution of link $i$ is obtained by evaluating the mixture model of \cite{OsoJing17} as defined in Equations~\eqref{eq:mix1} and \eqref{eq:mix2}.
Notice that steps 7(a) and 7(b) in the algorithm can be run simultaneously to further enhance the runtime. Additionally, for both steps 7(a) and 7(b) the calculations for each  link can be done in parallel (i.e., given their boundary conditions, links are updated independently). 
\begin{algorithm}[h]
	\caption{\label{algo1}Network model algorithm}
	\begin{enumerate}
		\item set values for the exogenous parameters $\hat{\varrho}_i, v_i, w_i, \ell_i$ and $\delta$ 
		\item compute $k_i^{fwd}=\lceil\frac{\ell_i}{\hat{\varrho}_iv_i\delta}\rceil$ and $k_i^{fwd}=\lceil\frac{\ell_i}{\hat{\varrho}_i|w_i|\delta}\rceil$ 
		\item compute $\hat{w}_i$ according to Eq. (37) of \cite{OsoJing17}
		\item set  initial conditions: $q_i^{in}(0)$, $q_i^{out}(0)$, $P^{UQ}(UQ_i(0))$, $P^{DQ}(DQ_i(0))$, $q^{UQ_i}(0)$, $q^{LLO_i}(0)$, $q^{LLI_i}(0)$, $q^{DQ_i}(0)$, $\lambda_i(0)$ and $\hat{\mu}_i(0)$.
		\item set $q_i^{in}(r)=0$ and $q_i^{out}(r)=0$ for $r<0$ 
		\item repeat the following for time intervals $k = 1,2,...$
		\begin{enumerate}
			\item for all links $i$, evaluate the univariate UQ model (i.e., run Algorithm 1 of \cite{OsoJing17}) with initial conditions $P^{UQ}(UQ_i(k-1))$, $\hat{\mu}_i(k-1)$ and $\lambda_i(k-1)$, this yields $P^{UQ}(UQ_i(k))$, $P^{UQ}(DQ_i(k))$ 
			\item for all links $i$, evaluate  the univariate DQ model (i.e., run Algorithm 2 of \cite{OsoJing17}) with initial conditions $P^{DQ}(DQ_i(k-1))$, $\hat{\mu}_i(k-1)$ and $\lambda_i(k-1)$, this yields  $P^{DQ}(DQ_i(k))$, $P^{DQ}(UQ_i(k))$
			\item calculate $P(DQ_i(k)>0,UQ_n(k)< \ell_n \ \forall n \in N)$ according to Eq. \eqref{eq:inclusive_exclusive}
			\item calculate $q_{i\rightarrow j}^{out}(k)$ according to Eq. \eqref{eq:node}
			\item calculate $q_i^{in}(k)$, $q_i^{out}(k)$ according to Eq. \eqref{eq:qIn_def1} and \eqref{eq:qOut_def1}, respectively.
			\item calculate  $\hat{\mu}_i(k)$ and $\lambda_i(k)$ according to Eq. \eqref{eq:effservice} and \eqref{eq:lambda}, respectively.
		\end{enumerate}
	\end{enumerate}
\end{algorithm}

\section{Model validation}\label{validation}	
In this section, we validate the proposed network model formulation. We consider the following network topologies: 
a three-link merge network (Section~\ref{merge_network}),
a three-link diverge network (Section~\ref{diverge_network})
and a network with 8 links with intricate traffic dynamics (Section~\ref{8-link}). The latter is the same network at that considered in Section 3 of~\citep{FloOso14_DTA}.
We carry out experiments with time-varying demand and varying levels of congestion.

The outputs of the analytical model  are compared to estimates obtained from a discrete-event  simulator of the stochastic link transmission model. This simulator has been used for validation in \citep{OsoFlo15_TS}.  The simulator  samples individual vehicles each with exact forward and backward lags, and it satisfies FIFO.
For each experiment, the simulated estimates are obtained from $10^6$ simulation replications. Estimates of 95\% confidence intervals were calculated, yet were barely visible. Hence, they are not displayed in the figures. 

For all networks, each link consists of a single lane. The lane parameters are common to all experiments, and are displayed in Table~\ref{tab:parameters}. The link lengths are fixed at $50$m (resp. space capacity $\ell=10$, forward lags $k^{fwd}\delta=5$ sec, and backward lags $k^{bwd}\delta=10$ sec). Each experiment starts with an empty network and runs for $600$ seconds.

\begin{table}
	\caption{Link Parameters}
	\begin{tabular*}{\hsize}{@{\extracolsep{\fill}}ll@{}}
		\hline
		Parameter & Value \\
		\hline
		$v$ & $0.01$ km/sec  \\
		$w$ & $-0.005$ km/sec \\
		$\hat{\varrho}$ & $200$ veh/km \\
		$\hat{q}$ & $2400$ veh/h = $0.67$ veh/sec \\
		$\delta$ & $0.1$ sec \\
		L ($\ell$, $k^{fwd}$, $k^{bwd}$) & $50$ m ($10$,  $50$, $100$) \\
		\hline
	\end{tabular*}	
	\label{tab:parameters}
\end{table}

\subsection{Three-link merge network}\label{merge_network}

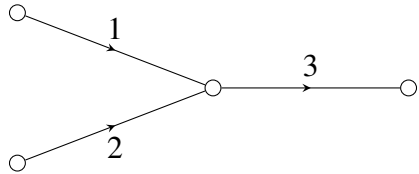
\begin{figure}
	\centering
	\begin{tikzpicture}[x=1.3cm, y=1cm,
	every edge/.style={
		draw,
		postaction={decorate,
			decoration={markings,mark=at position 0.5 with {\arrow{stealth}}}}
	}
	]
	\vertex (1) at (0,2) {};  
	\vertex (2) at (0,0) {};
	\vertex (3) at (2,1) {};
	\vertex (4) at (4,1) {};
	\path
	(1) edge node[above]{$1$} (3)
	(2) edge node[below]{$2$} (3)
	(3) edge node[above]{$3$} (4)
	;  
	\end{tikzpicture}
	\caption{Merge network of three links}
	\label{3_network1}
\end{figure}

The topology of the three-link merge networks is  shown in Figure~\ref{3_network1}. Vehicles enter the network by joining link 1 or 2, and they leave the network through link 3. The service rates $\mu_i$ for link $i$ ($i=1,2,3$) are exogenous, and the arrival rates $\lambda_i$ to links 1 and 2 ($i \in \{1,2\}$) are exogenous, whereas $\lambda_3$ is endogenous. We conduct 2 experiments with time-varying demand for link 1 and constant demand for link 2. The experimental design is detailed in Table~\ref{tab:2}.
For each experiment, the arrival rate is 0.15 veh/sec during time [0,200] seconds, 0.25 veh/sec during time [200,400] seconds and 0.05 veh/sec during time [400,600] seconds. 

\begin{sidewaysfigure}
	\minipage{1\textwidth}
	\includegraphics[width=\linewidth]{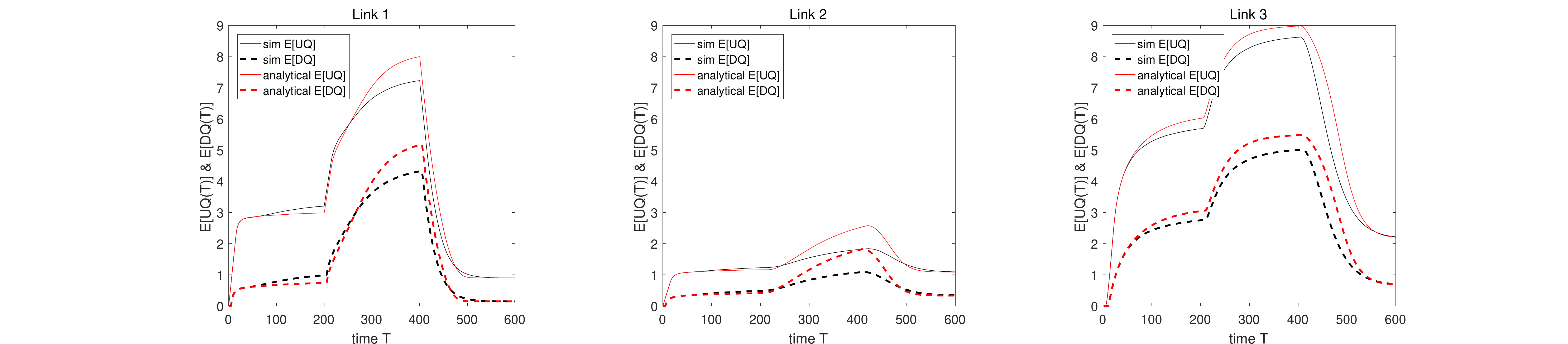}
	\subcaption{Experiment 1}
	\label{fig:5}
	\endminipage\hfill
	\minipage{1\textwidth}
	\includegraphics[width=\linewidth]{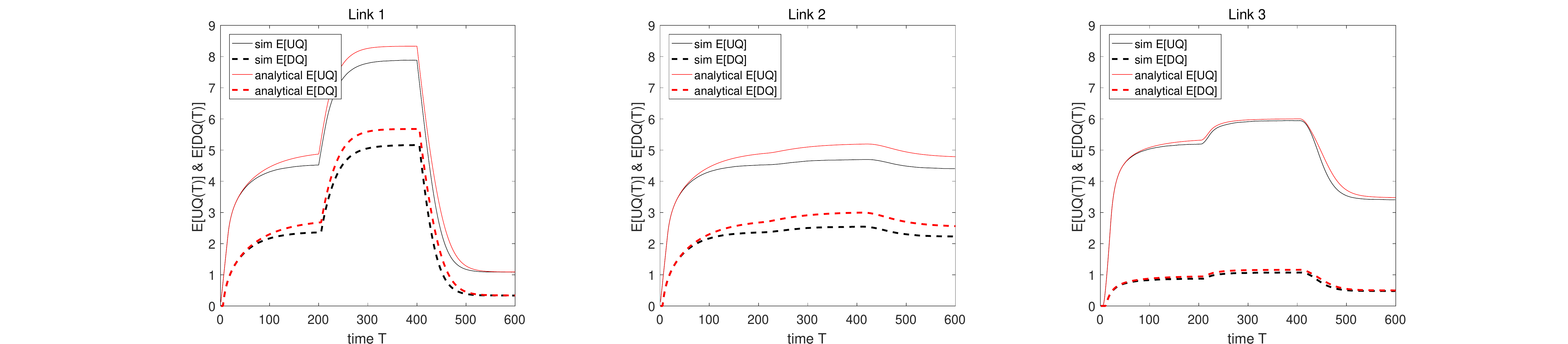}
	\subcaption{Experiment 2}
	\label{fig:6}
	\endminipage\hfill
	\caption{$E[UQ(T)]$ and $E[DQ(T)]$}
	\label{fig:3-link merge}
\end{sidewaysfigure}

For experiment 1, the congestion level of link 1 is expected to increase and then decrease, while link 2 is expected to be uncongested. During the first 400 seconds a spillback from link 3 to links 1 and 2 is expected. For experiment 2, link 1 is expected to be congested for the first 400 seconds and uncongested thereafter. Link 2 is expected to be congested, and link 3 is expected to be uncongested all the time. Figure~\ref{fig:5} (resp. Fig.~\ref{fig:6}) displays $E[UQ(T)]$ and $E[DQ(T)]$ as a function of time $T$ for experiment 1 (resp. experiment 2).

\begin{table}
	\caption{Experimental design}
	\begin{tabular*}{\hsize}{@{\extracolsep{\fill}}lllllll@{}}
		\hline
		Experiment &  network type &             $\lambda_1$                & $\mu_1$ & $\lambda_2$ & $\mu_2$ & $\mu_3$ \\
		\hline
		1      & merge &$ 0.15 \rightarrow 0.25\rightarrow 0.05$ &  $0.4$  &   $0.05$    &  $0.2$  &  $0.2$  \\
		2      & merge &$ 0.15 \rightarrow 0.25\rightarrow 0.05$ &  $0.2$  &   $0.15$    &  $0.2$  &  $0.6$  \\
		\hline
		Experiment & network type & $\lambda_1$ & $\mu_1$ & $\mu_2$ & $\mu_3$ & $p_{12}$ \\
		\hline
		3     & diverge & $ 0.2 \rightarrow 0.4 \rightarrow 0.1$ &  $0.3$  & $0.05$  &  $0.4$  &  $0.2$   \\
		4     & diverge & $ 0.2 \rightarrow 0.4 \rightarrow 0.1$ &  $0.6$  &  $0.4$  &  $0.4$  &  $0.5$   \\
		\hline
	\end{tabular*}
	\label{tab:2}
\end{table}

The plots in columns 1, 2, and 3 display, respectively, the results for links 1, 2 and 3. 
For link 1,  there is, as expected, an increase in expectation during time [200, 400] seconds, which is caused by both the increase in demand and the occurrence of spillback from link 3. Then, there is a sharp decrease after time $T=400$ seconds caused by the decrease in demand and the spillback dissipation. 

For link 2 (middle plot), there is an increase in expectation during time [200, 400] seconds.  This increase is purely caused by the spillback from link 3.
For link 3, there is an increase in expectation during time [200, 400] seconds and a decrease after time $T=400$ seconds caused by the changes in demand of link 1. During time [200, 400] seconds, $E[UQ(T)]$ almost reaches the space capacity and causes a significant spillback effect from link 3 to links 1 and 2. For all three plots of Figure~\ref{fig:5}, the  proposed analytical model yields approximations which accurately capture the trends of the simulated estimates.
However, the  proposed model  overestimates the spillback of link 3, and its effect on links 1 and 2.

Figure \ref{fig:6} displays the results of experiment 2. It has the same  structure as Figure~\ref{fig:5}. The increase and decrease in demand of link 1 causes the increase and decrease in expectation for both links 1 and 3. For link 2 with constant demand, there is a subtle increase and decrease in expectation. Even this subtle change is captured by the proposed model. Overall, the proposed model captures  the temporal variations of these performance metrics accurately.

\subsection{Three-link diverge network}\label{diverge_network}
\begin{figure}
	\centering
	\begin{tikzpicture}[x=1.3cm, y=1cm,
	every edge/.style={
		draw,
		postaction={decorate,
			decoration={markings,mark=at position 0.5 with {\arrow{stealth}}}}
	}
	]
	\vertex (1) at (0,1) {};  
	\vertex (2) at (2,1) {};
	\vertex (3) at (4,2) {};
	\vertex (4) at (4,0) {};
	\path
	(1) edge node[above]{$1$} (2)
	(2) edge node[above]{$2$} (3)
	(2) edge node[below]{$3$} (4)
	;  
	\end{tikzpicture}
	\caption{Diverge network of three links}
	\label{3_network2}
\end{figure}
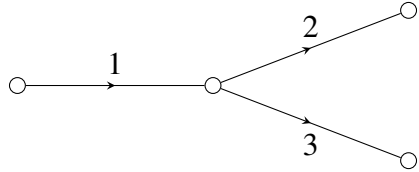

The  topology of the three-link diverge network  is shown in Figure~\ref{3_network2}. Vehicles enter the network by joining link 1 at a rate $\lambda_1$. From link 1, they continue to either  link 2 with turning probability $p_{12}$ or to link 3 with probability $p_{13}$ (i.e., $p_{13} = 1 - p_{12}$).
The two experiments that we conduct consider
time-varying demand for link 1 and different turning probabilities. The experimental design is presented in Table~\ref{tab:2}. For each experiment, the arrival rate is 0.2 veh/sec during time [0,200] seconds, 0.4 veh/sec during time [200,400] seconds and 0.1 veh/sec during time [400,600] seconds.

For both experiments 3 and 4, the congestion level of all links is expected to increase and then decrease over time. For experiment 3, the congestion levels of downstream links 2 and 3 will  differ. For experiment 4, both downstream links have identical configuration, and the vehicles have equal probability of choosing either link, hence  we expect the congestion levels of links 2 and 3 to be identical. 

\begin{sidewaysfigure}
	\minipage{1\textwidth}
	\includegraphics[width=\linewidth]{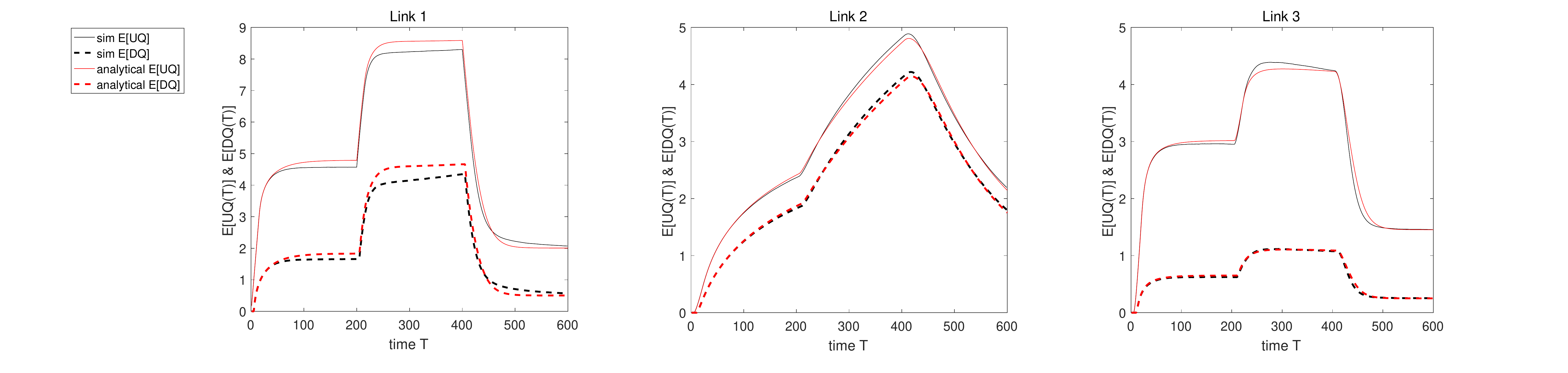}
	\subcaption{Experiment 3}
	\label{fig:7}
	\endminipage\hfill
	\minipage{1\textwidth}
	\includegraphics[width=\linewidth]{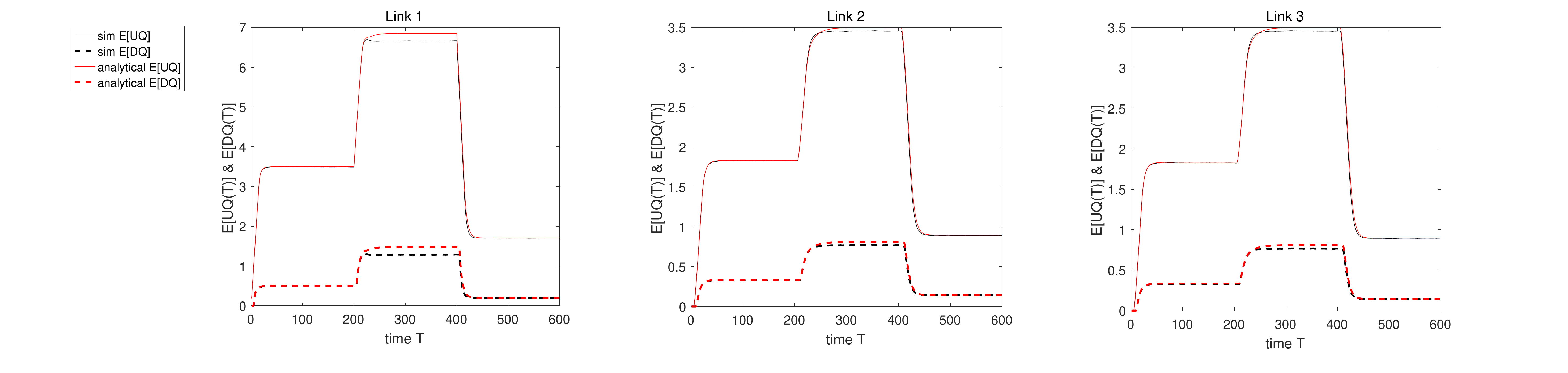}
	\subcaption{Experiment 4}
	\label{fig:8}
	\endminipage\hfill
	\caption{$E[UQ(T)]$ and $E[DQ(T)]$}
	\label{fig:3-link diverge}
\end{sidewaysfigure}

Figure~\ref{fig:7} shows the results for experiment 3. The graphs from left to right display $E[UQ(T)]$ and $E[DQ(T)]$ for links 1, 2, and 3, respectively. As a first impression, the deviations between analytical and simulated expectations are small. The shapes of expectation curves for links 2 and 3 are different. The sharp increase in expectation during the second 200 seconds and the sharp decrease in expectation after $400$ seconds for all three links are well approximated by the proposed model.
Figure~\ref{fig:8} shows the results for experiment 4. The proposed model yields identical expectation curves for links 2 and 3, as expected.  The analytical approximate expectations match the simulated estimates. The proposed model accurately captures the temporal variations, in both shape and   magnitude, of these performance metrics.

\subsection{Eight link network}\label{8-link}
\begin{figure}
	\centering
	\begin{tikzpicture}[x=1.3cm, y=1cm,
	every edge/.style={
		draw,
		postaction={decorate,
			decoration={markings,mark=at position 0.5 with {\arrow{stealth}}}}
	}
	]
	\vertex (1) at (0,0) {};  
	\vertex (2) at (2,0) {};
	\vertex (3) at (4,0) {};
	\vertex (4) at (6,0) {};
	\vertex (5) at (0,2) {};  
	\vertex (6) at (2,2) {};
	\vertex (7) at (4,2) {};
	\vertex (8) at (6,2) {};
	\path
	(1) edge node[above]{$1$} (2)
	(2) edge node[above]{$6$} (3)
	(3) edge node[above]{$4$} (4)
	(8) edge node[above]{$2$} (7)
	(7) edge node[above]{$8$} (6)
	(6) edge node[above]{$3$} (5)
	(2) edge node[left]{$5$} (6)
	(7) edge node[right]{$7$} (3)
	;  
	\end{tikzpicture}
	\caption{Toy network of eight links}
	\label{8_network}
\end{figure}
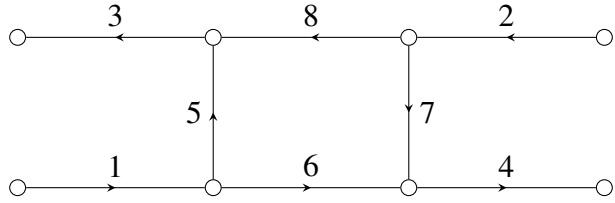
\begin{figure}
	\includegraphics[width=\linewidth]{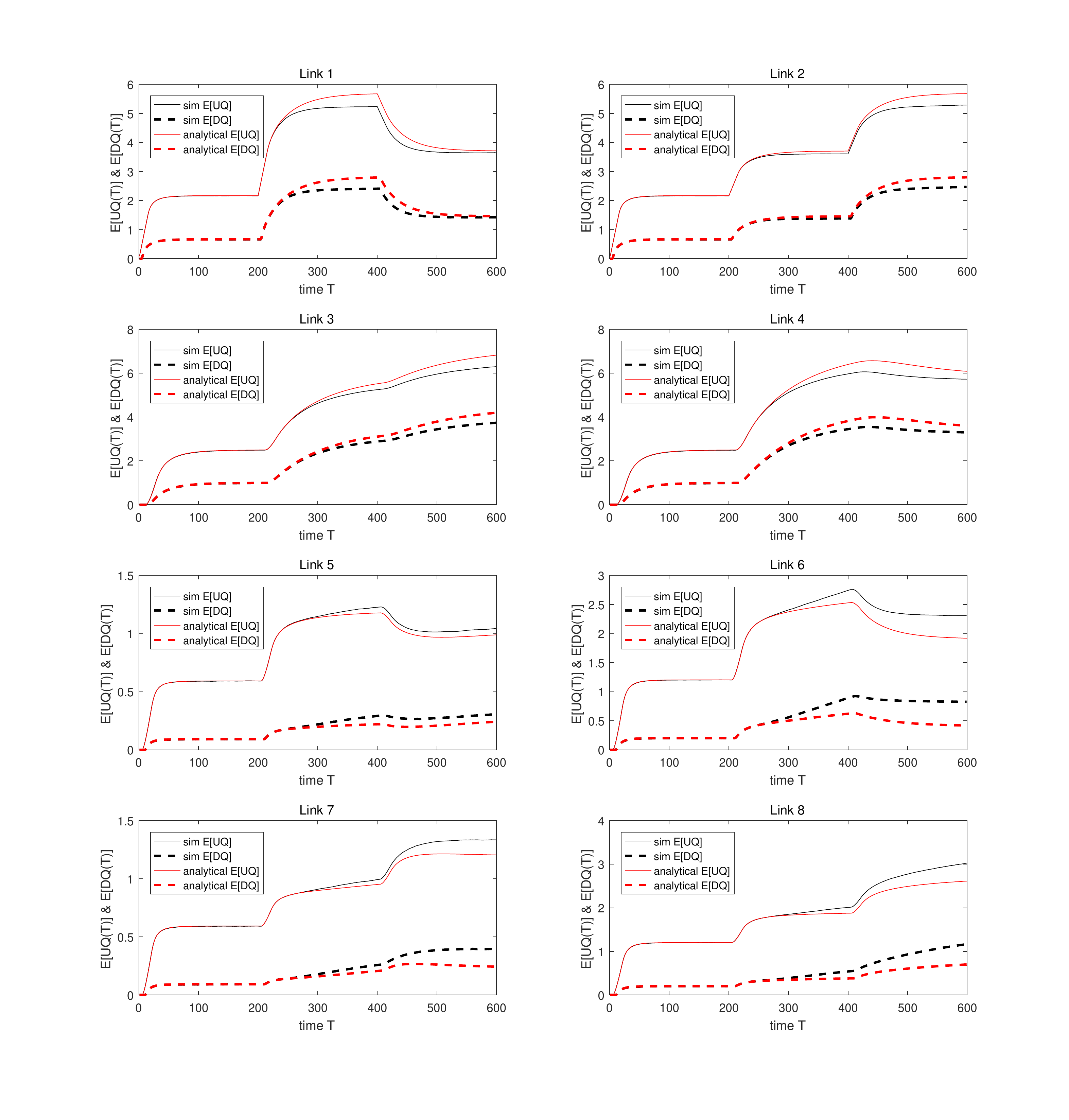} 
	\caption{$E[UQ(T)]$ and $E[DQ(T)]$ for the experiment on toy network}
	\label{fig:E}
\end{figure}

Finally, we  consider an eight link network with intricate traffic dynamics.
The network topology is given in Figure~\ref{8_network}. This network configuration is the same as in \cite{FloOso14_DTA}. Vehicles enter the network by joining links 1 or 2. They can leave the network through links 3 and 4. Upon entrance to link 1 (resp. 2), vehicles  can either go straight to link 6 (resp. 8) and leave the network through link 4 (resp. 3) or initiate a U-turn to link 5 (resp. 7) and leave the network through link 3 (resp. 4). The concrete parameters used for this  network are as follows. Vehicles can enter the network through links 1 or 2.  For link 1, vehicles enter at a rate of 0.1 veh/sec during [0,200] seconds, 0.2 veh/sec during [200, 400] seconds and 0.15 veh/sec during [400,600] seconds. For link 2, vehicles enter at a rate of 0.1 veh/sec during [0,200] seconds, 0.15 veh/sec during [200, 400] seconds and 0.2 veh/sec during [400,600] seconds. 
They continue straight with probability $p_{16}=p_{28}=2/3$ and perform a U-turn with probability $p_{15}=p_{27}=1/3$. The service rates for link 1 and 2 are $\mu_1=\mu_2=0.25$ veh/sec and the service rates for the links within the network are $\mu_5=\mu_6=\mu_7=\mu_8=0.4$ veh/sec, which is on average sufficient to serve the  demand. The service rates for exit links 3 and 4 are lower at $\mu_3=\mu_4=0.2$ veh/sec. 

The difference in time-varying demands at the two source links 1 and 2 leads to intricate traffic patterns. The congestion patterns that we expect to observe based on this experimental design are as follows.  Link 1 experiences step-changes from uncongested to congested, and then again to uncongested conditions. Link 2, on the other hand, experiences step-changes steadily from uncongested to congested conditions. For the first 200 seconds, the exit link 3 (resp. 4) is uncongested, hence the congestion level of its upstream links  5 and 8 (resp. 6 and 7) are determined by the outflow from their respective upstream links 1 (resp. 2).
After 200 seconds, travel demand increases. Thus, the occurrence of spillback on the exit link 3 (resp. 4) increases, this impacts the congestion levels of its upstream links 5 and 8 (resp. 6 and 7).

Each plot of Figure~\ref{fig:E} shows the evolution of $E[UQ(T)]$ and $E[DQ(T)]$ for one of the eight links in the network. The line styles and colors are the same as for the previous figures.
The temporal variations  of the analytical approximations mimic well the simulated variations.  However, it tends to underestimate  the congestion levels of links 6 and 8. 

In summary, the proposed model captures the temporal evolution in expectations for this network with intricate traffic patterns. Additionally, it does so with a significant gain in computational efficiency. For instance, since each link of this network has a space capacity of 10, the state space of the  joint network distribution has dimension $11^8$. The proposed model overcomes the need to model the entire joint distribution. Instead, the memory   required for the proposed model is $2\times 11 \times 8=176$ numbers. This represents a significant gain in computational efficiency, and enables its use for large-scale network analysis and optimization.

\section{Case study}\label{casestudy}
\begin{figure}[htp]
	\minipage{1\textwidth}
	\centering
	\includegraphics[width=0.8\linewidth]{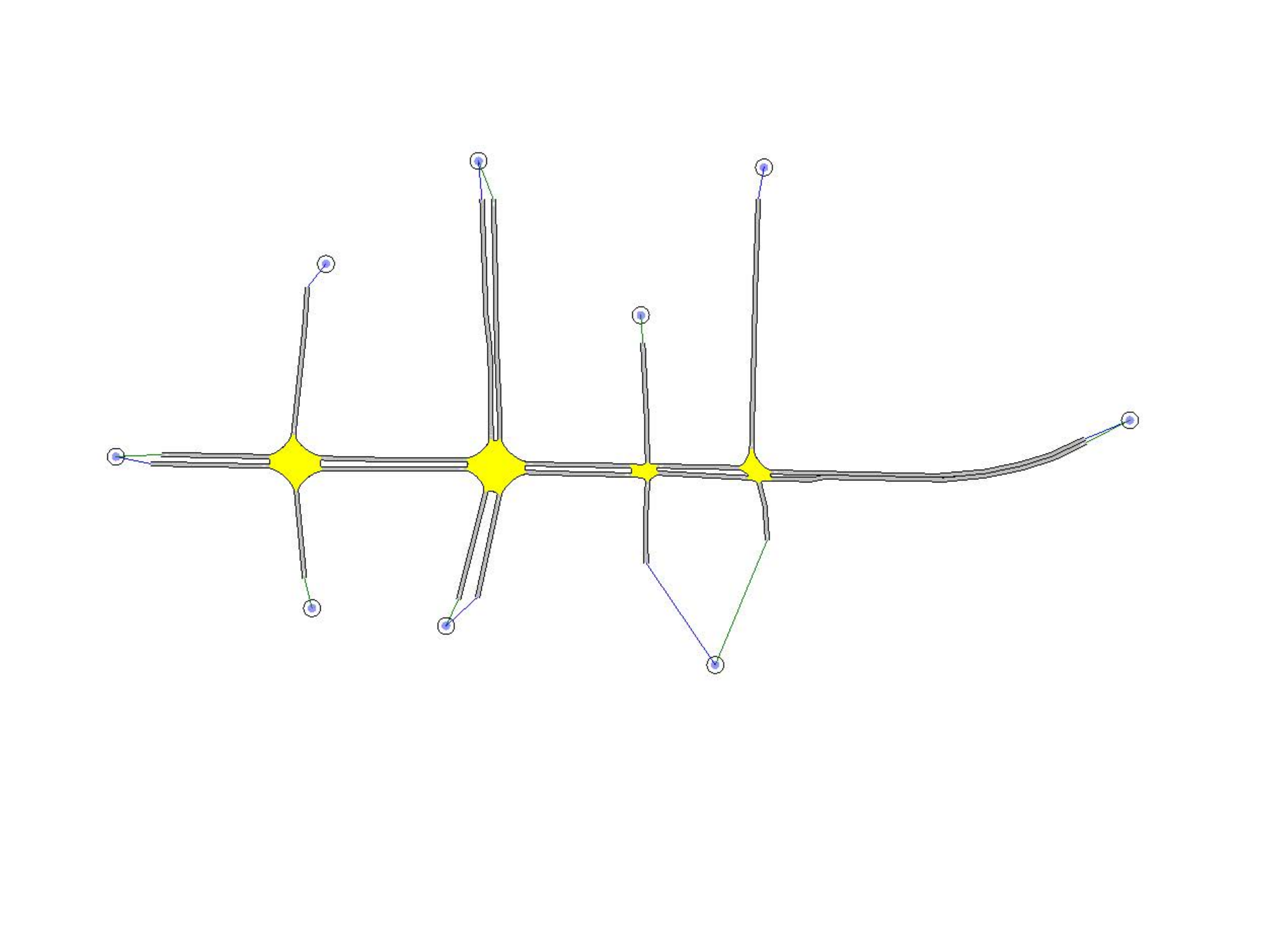}
	\caption{Network topology}\label{fig:roadnet}
	\endminipage\hfill
	\minipage{1\textwidth}
	\centering
	\includegraphics[width=0.8\linewidth]{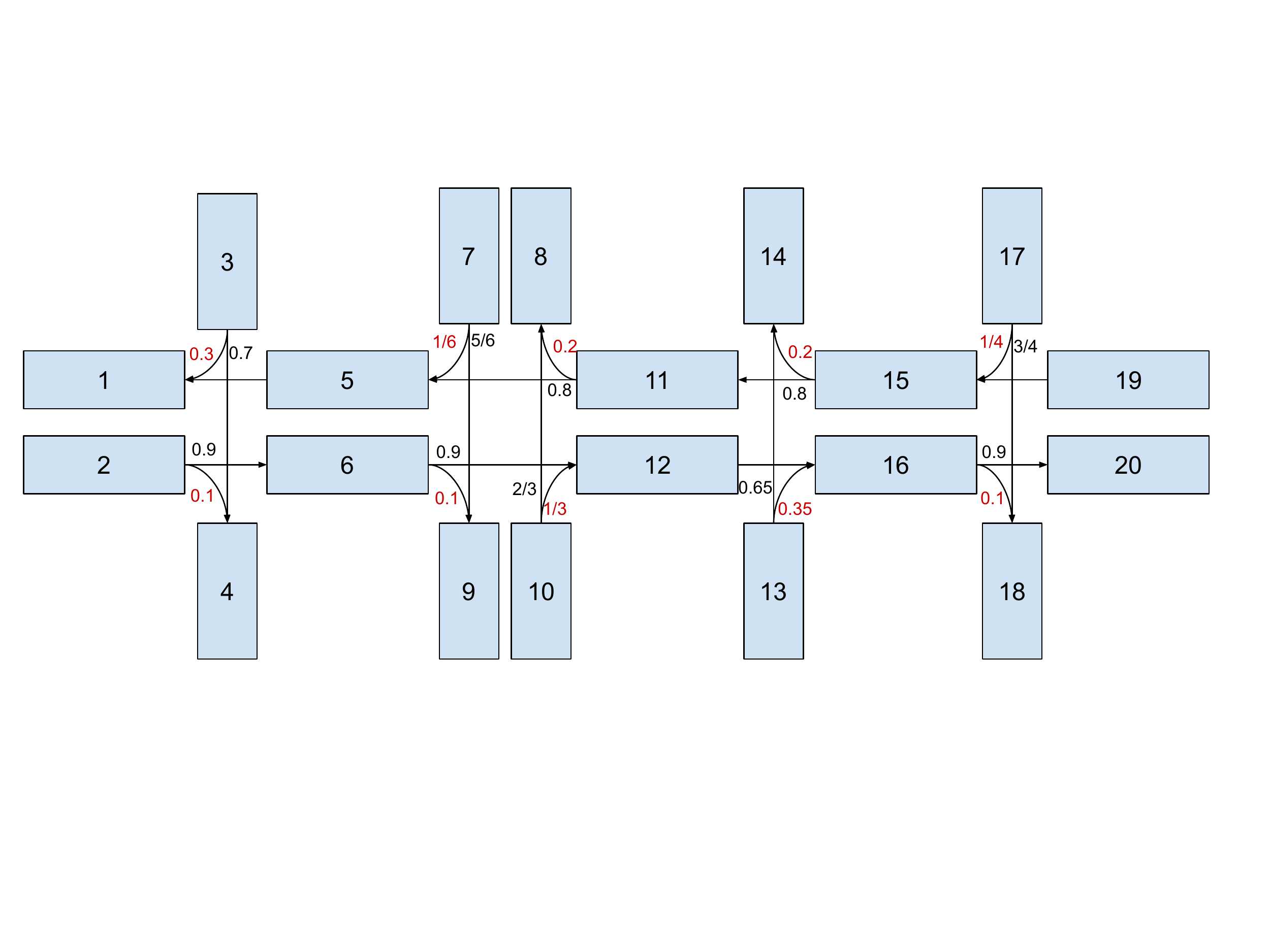}
	\caption{Network model}\label{fig:queuenet}
	\endminipage\hfill
\end{figure}

In this section, the proposed network model is used to address an optimization problem. 
The purpose of this section is to investigate the added value of the proposed network model, and more specifically of the node model, for traffic control.
We consider a fixed-time signal control problem for the network shown in Figure~\ref{fig:roadnet}.
The representation of the road network as a single-lane network is presented in Figure~\ref{fig:queuenet}.
Section~\ref{problem} formulates the problem and describes the case study. Section~\ref{OptResults} presents the numerical results. 

We address the problem with the proposed network model. 
We also solve it with the mixture model (described in Section \ref{link_model}) and with an exogenous node model (described in Section 4.1 of \citep{OsoJing17}). 
Hereafter, we call the latter approach the marginal model. In the marginal model, the total demand to each link is exogenous. Hence, each link can be modeled independently. That is, the marginal model accounts for the within-link temporal variations of congestion, but does not capture any between-link (i.e., across-node) interactions.
In other words, both models (proposed and marginal) consider the same link model. The proposed model has an endogenous node model (formulated in Section \ref{node_model}) which captures the between-link interactions, while the marginal model omits between-link interactions. 
We use both models to address the optimization problem and compare their performances. This comparison will indicate the added value of accounting for between-link interactions in traffic control.

Additionally, the deterministic intelligent link transmission model (ILTM) \citep{HimpeWillem2016IAfR}, which is a complete network loading model that embeds the well-established node model of \citet{Tampere11}, is applied to address the same problem. The comparison between the performance of the ILTM and the proposed model will indicate the added value of accounting for stochasticity in traffic control.

\subsection{Network configuration and problem formulation}\label{problem}

We consider a synthetic network, which is displayed in Figure~\ref{fig:roadnet}. 
The network consists of $20$ single-lane roads and $4$ intersections, each with $2$ endogenous signal phases. The circles in Figure~\ref{fig:roadnet} are where the vehicles can enter and leave the network. 
We consider a fixed-time signal control problem. We refer the reader to Appendix A of \citep{OsorioPhD} for a review of traffic signal control terminology and formulations. 
Fixed-time signal plans are commonly used for networks with sparse or unreliable real-time data, they are also used by major cities, such as New York City \cite{osoChen15_qbbMTMTechRep}, with high and uniformly distributed congestion levels.

We consider a fixed-time signal control problem for a $30$ min period. The signal plans of the $4$ intersections are jointly determined. The decision variables are the green splits of the phases of the different intersections. All other traditional control variables (e.g., cycle times, offsets, stage structure) are assumed fixed. There are $2$ endogenous signal phases per intersection. Hence, the dimension of the decision variable vector is $8$.

To formulate the signal control problem, we introduce the following notation:
\[
\begin{tabular*}{0.5\textwidth}{@{\extracolsep{\fill} }ll}
$b_d$ & ratio of available cycle time to total cycle time for intersection $d$; \\
$x$ & vector of green splits;\\
$x(j)$ & green split of signal phase $j$; \\
$x_{LB}$ & vector of lower bounds for green splits; \\
$\mathcal{D}$ & set of intersection indices; \\
$\mathcal{P}_D(d)$ & set of endogenous signal phase indices of intersection $d$; \\
$\mathcal{L}$ & set of all links;\\
$\tilde{T}$ & the length of the time interval of interest [in min]; \\
$|\mathcal{L}|$ & the number of links, i.e., cardinality of $\mathcal{L}$. \\
\end{tabular*}
\label{terminology}
\]
The problem is formulated as follows:
\begin{eqnarray}
\min_{x} f(x)=\frac{1}{\tilde{T}|\mathcal{L}|}\sum_{i\in \mathcal{L}}\sum_{\hat{t}=1}^{\tilde{T}}E[DQ_i(\hat{t};x)]
\label{eq:obj}
\end{eqnarray}
subject to
\begin{eqnarray}
\sum_{j\in\mathcal{P}_D(d)}x(j)&=&b_d, \quad \forall d \in \mathcal{D} \label{constr1}\\
x&\geq& x_{LB}. \label{constr2}
\end{eqnarray}
The decision vector, $x$, denotes the green splits of the signal controlled links. Constraint~\eqref{constr1} ensures that for each intersection, the sum of green splits equals the available cycle time. Constraint~\eqref{constr2} ensures lower bounds for the green splits. The objective function represents the average, over time and over links, of the expectation of $DQ$. $E[DQ_i(\hat{t};x)]$ denotes the expectation of $DQ$  of link $i$ at integer time $\hat{t}$ (in minutes) under signal plan $x$. It is calculated by evaluating either the proposed network model or the marginal link model.

For the deterministic ILTM, the objective function is the average, over time and over the network lanes, number of vehicles on the link, i.e.,
\begin{equation}
f(x)=\frac{1}{\tilde{T}|\mathcal{L}|}\sum_{i\in \mathcal{L}}\sum_{\hat{t}=1}^{\tilde{T}}\left(c^{up}_i(\hat{t};x)-c^{down}_i(\hat{t};x)\right). \label{eq:DetObj1}
\end{equation}
where $c^{up}_i(\hat{t};x)$ (resp. $c^{down}_i(\hat{t};x)$) is the cumulative number of vehicles that cross the upstream (resp. downstream) end of link $i$ up until time $\hat{t}$ under signal plan $x$, which is the output metric produced by the ILTM.

Note that the decision variables (i.e., the green splits) are related to the link parameters as follows. Given a signal plan, the service rate, $\mu_i$, of signalized link $i$  is determined by the following equation.
\begin{equation}
\mu_i-\sum_{j\in\mathcal{P}_{I}(i)}x(j)s=e_is
\end{equation}
where $s$ represents the saturation flow, $e_i$ represents the ratio of fixed green time to cycle time and
$\mathcal{P}_{I}(i)$ represents the set of endogenous signal phases of link $i$. The ILTM also considers an average outflow capacity as suggested by the developers. This is because the explicit implementation of traffic cycles comes at a cost of more memory use (which can be prohibitive in Matlab) and higher computation times.

Table~\ref{tab:parameterLan} displays the values of the main exogenous model parameters. Note that there are two sets of demand scenarios in Section 4 of \citep{OsorioYamani_forthcoming}, high demand and medium demand, we are going to carry out the optimization for both demand scenarios. The optimization problems are solved using the \textit{Active-set} algorithm of the \textit{fmincon} routine of \citep{MATLAB:2016}.

\begin{table}
	\caption{Parameters for the case study}
	\begin{tabular*}{\hsize}{@{\extracolsep{\fill}}ll@{}}
		\hline
		Parameter & Value \\
		\hline
		$\tilde{T}$ & $15$ min\\
		$|\mathcal{L}|$ & $20$ links\\
		$\delta$ & $0.1$ sec\\
		$x_{LB}$ & $4$ sec\\
		$v$ & $50$ km/h  \\
		$w$ & $-15$ km/h \\
		$\hat{\varrho}$ & $200$ veh/km \\
		$s$ & $1800$ veh/h\\
		$\mu$ & varies by signal plans \\
		$\ell$, $\gamma$ & exogenous values from section 4 of \cite{OsorioYamani_forthcoming} \\
		$p_{ij}$ & turning factions is shown in Figure \ref{fig:queuenet}\\
		$k^{fwd}$ & $k^{fwd}=\lceil\frac{\ell}{\hat{\varrho}v}\rceil$\\
		$k^{bwd}$ & $k^{fwd}=\lceil\frac{\ell}{\hat{\varrho}|w|}\rceil$\\
		\hline
	\end{tabular*}	
	\label{tab:parameterLan}
\end{table}

\subsection{Numerical analysis}\label{OptResults}

To initialize the algorithm, we randomly and uniformly draw a point from the  feasible space (Equations \eqref{constr1}-\eqref{constr2}). We use the code of
\citep{Stafford2006} to perform this uniform sampling. 
We initialize the proposed model, the marginal model and the ILTM with the same random initial point.
We then solve Problem~\eqref{eq:obj}-\eqref{constr2} with each of the models. Each model yields an optimal signal plan. 
To evaluate the performance of a given signal plan, we use a microscopic traffic simulation model implemented with the Aimsun simulator \citep{TSSAM}. 
For a given signal plan, we embed it within the simulator and run $50$ simulation replications. Each simulation replication simulate a $30$ min period. Each replication yields one observation of the main performance measures: average link queue-length and average trip travel time. We then compare the cumulative distribution functions (cdf's) obtained from the $50$ replications of these performance measures.

\begin{figure}
	\vspace{-3cm}
	\minipage{1\textwidth}
	\centering
	\includegraphics[width=0.9\linewidth]{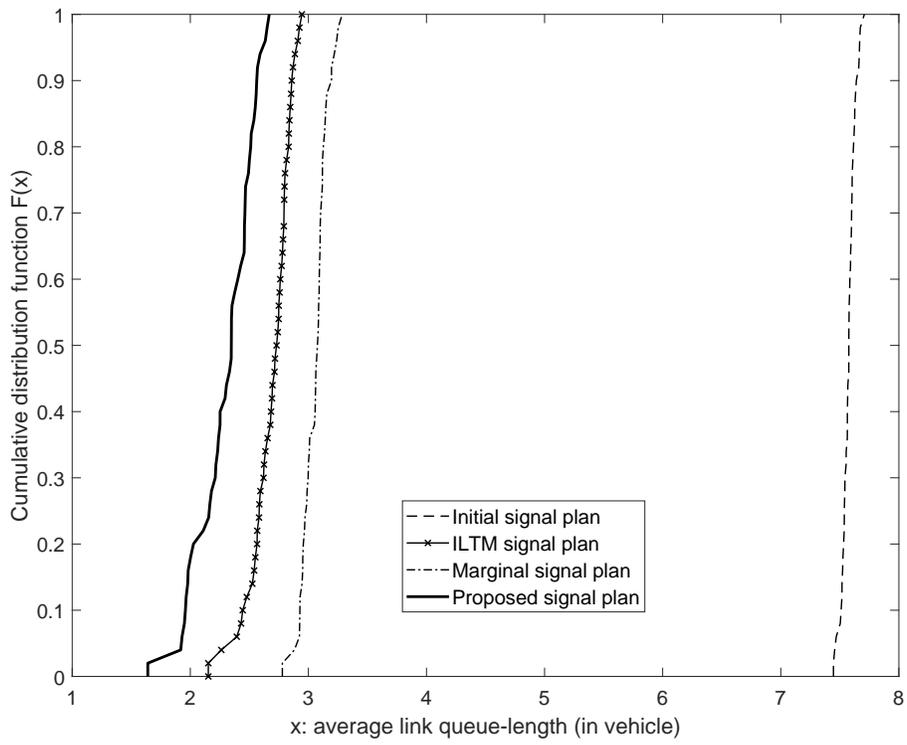}
	\subcaption{High demand scenario}\label{fig:meanQ_high}
	\endminipage\hfill
	\minipage{1\textwidth}
	\centering
	\includegraphics[width=0.9\linewidth]{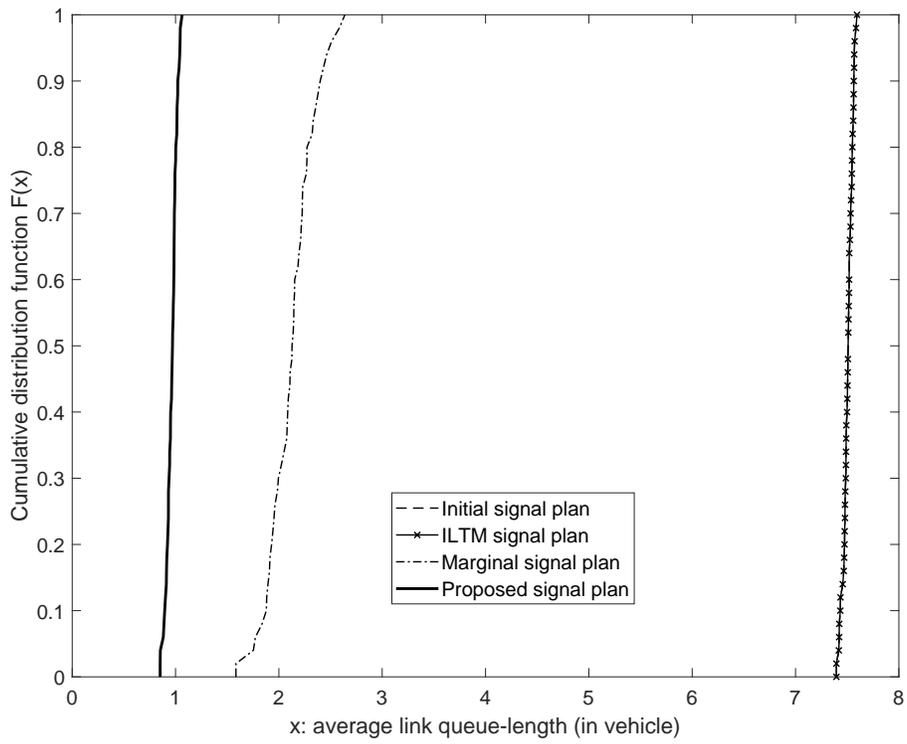}
	\subcaption{Medium demand scenario}\label{fig:meanQ_med}
	\endminipage\hfill
	\caption{Cumulative distribution functions of the average, over all lanes in the network, link queue-length in vehicles.}
	\label{fig:meanQ}
\end{figure}

Figures~\ref{fig:meanQ_high} and~\ref{fig:meanQ_med} consider the average link queue-length for high and medium demand scenarios, respectively. 
For each demand scenario, it displays four cdf curves: the initial signal plan (dashed line), the signal plans derived by the proposed network model (solid line), by the marginal model (dot-dashed line) and by the ILTM (crossed line).
The average link queue-length is an average obtained over time and over all links in the network. The $x$-axis displays the average link queue-length. For given $x$ value, the $y$-axis displays the proportion of simulation replications (out of $50$) that have average queue-lengths smaller than $x$. Therefore, the more a cdf curve is to the left, the better the performance of the corresponding signal plan. 

For high demand scenario, all models yield signal plans with improved performance compared to the initial plan.
Additionally, the cdf curve of the proposed model is to the left of both cdf curves from the ILTM and the marginal model. This indicates the signal plan derived by the proposed model outperforms that derived by the ILTM and by the marginal model. 
The signal plan of the proposed model yields a $25\%$ improvement in the mean average link queue-length compared to the signal plan of the marginal model ($2.31$ versus $3.06$) and a $14\%$ improvement compared to the ILTM ($2.31$ versus $2.69$). 

For medium demand scenarios, the ILTM does not detect objective function improvement direction at the initial point and hence the optimization stopped with the initial point randomly generated. Therefore, in Figure~\ref{fig:meanQ_med}, the curve of the ILTM completely overlaps with that of the initial signal plan. This is because the ILTM calculates the deterministic average inflow and outflow capacity of the link over time and hence the model is not sensitive to the change of outflow capacity as long as it is large enough to allow all inflow to pass through. On the other hand, the stochastic models are sensitive to changes in outflow capacity (i.e., $\mu$), especially when traffic intensity is near 1. The cdf curves of both the proposed and the marginal model are to the left of that of the ILTM. This indicates the signal plans derived by the proposed and marginal models outperform that derived by the ILTM. Additionally, the proposed model outperforms the marginal model and yields a $55\%$ improvement in the mean average link queue-length ($0.97$ versus $2.13$).

\begin{figure}
	\vspace{-3cm}
	\minipage{1\textwidth}
	\centering
	\includegraphics[width=0.9\linewidth]{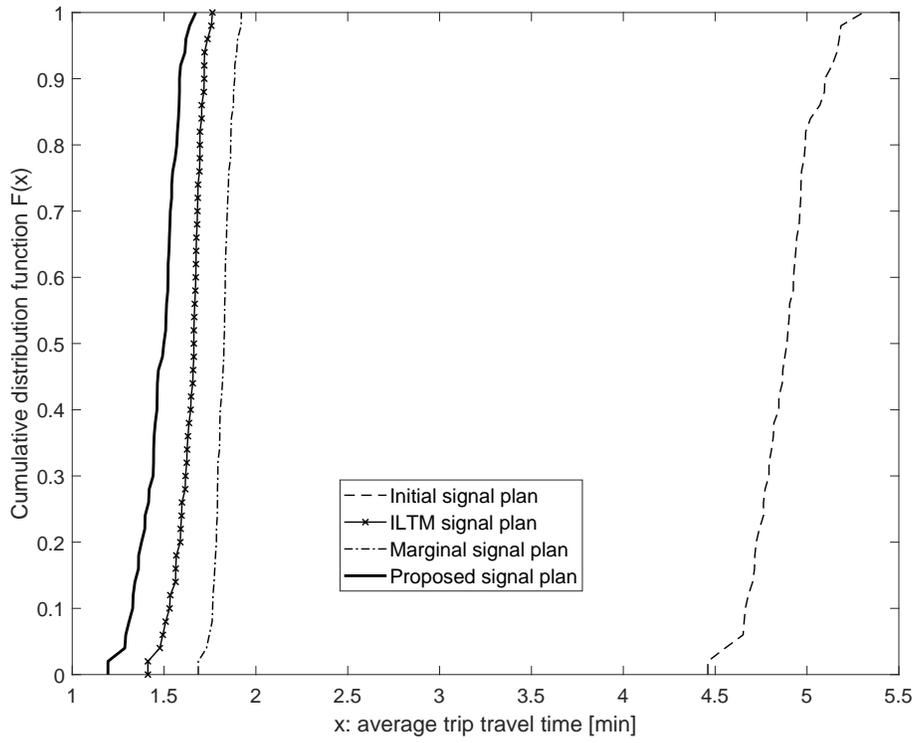}
	\subcaption{High demand scenario}\label{fig:meanTTT_high}
	\endminipage\hfill
	\minipage{1\textwidth}
	\centering
	\includegraphics[width=0.9\linewidth]{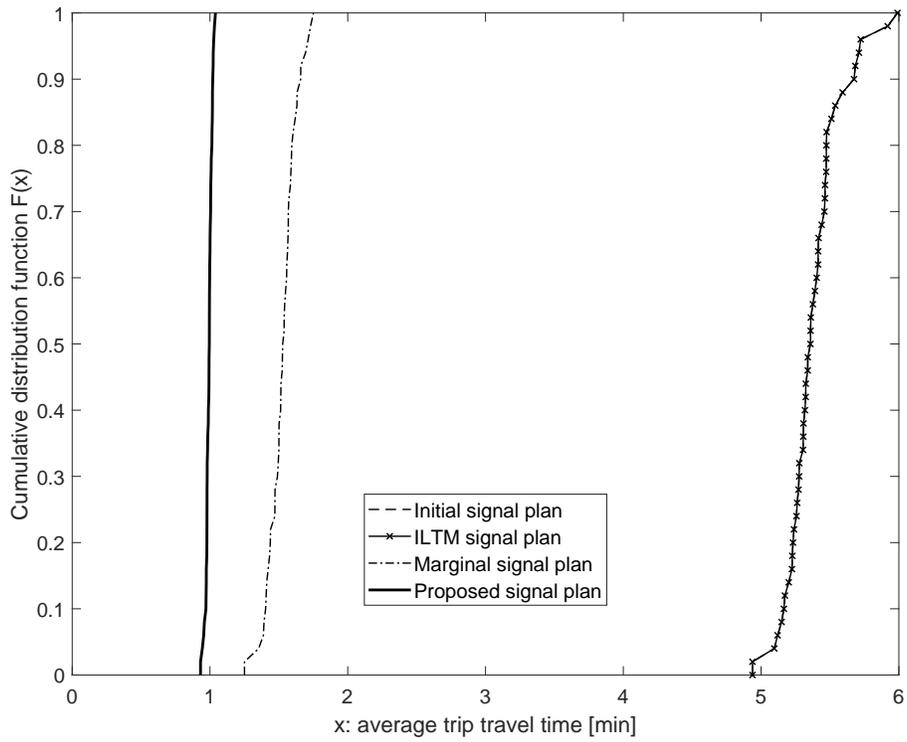}
	\subcaption{Medium demand scenario}\label{fig:meanTTT_med}
	\endminipage\hfill
	\caption{Cumulative distribution functions of the average trip travel time in minutes.}
	\label{fig:meanTTT}
\end{figure}

Figures~\ref{fig:meanTTT_high} and~\ref{fig:meanTTT_med} compare the performance of the four signal plans considering the average trip travel times (in minutes) for high and medium demand scenarios, respectively. The $x$-axis shows average trip travel time. the $y$-axis displays the proportion of simulation replications (out of $50$) that have average trip travel times smaller than $x$. Figure~\ref{fig:meanTTT} has a similar layout as Figure~\ref{fig:meanQ}.
The same conclusions as for Figure~\ref{fig:meanQ} hold. All models yield signal plans no worse than the initial signal plan in both demand scenarios.
The proposed model yields signal plans that outperform those derived by the ILTM and the plan derived by the marginal model in both demand scenarios.
The signal plan of the proposed model yields an $35\%$ improvement in the mean average trip travel time compared to the signal plan of the marginal model ($0.99$ minutes versus $1.53$ minutes) in medium demand scenario and $19\%$ improvement in high demand scenario ($1.48$ minutes versus $1.82$ minutes). The ILTM performances better than the marginal model in high demand scenario but not in medium demand scenario.

This case study shows the proposed network model outperforms the marginal model and the ILTM. It demonstrates the added value of the proposed node model to address the between-link interactions. Moreover, by considering stochasticity, the proposed network model, compared to the ILTM, is able to identify signal plans with better and more robust performance in various demand level.


\section{Conclusion}\label{conclusion}
This paper formulates an analytical stochastic node model. It formulates a full network model by coupling the node model with the link model of \citep{OsoJing17}.
The proposed network model is scalable and tractable.
In particular, the node model is not based on the joint modeling of all upstream and downstream links adjacent to a node. Instead the joint probabilities are approximated from the univariate distributions of the upstream and downstream links. 
The node model is formulated for nodes with an arbitrary configuration, it satisfies flow conservation properties.
The proposed network model is validated versus stochastic simulation results, using a simulator of the stochastic link transmission model. 
We consider a set of small networks with intricate time-varying congestion patterns. The boundary conditions of every link in the network are well approximated by the proposed network model. 
The proposed model is used to address a signal control problem for a synthetic network with different demand level. We compare the performance of the proposed approach to that of a model with the same link model but without a node model (i.e., between-link interactions are not captured) as well as a deterministic ILTM with well-established deterministic node model. The proposed network model yields signal plans that outperform the marginal model as well as the deterministic ILTM.
It shows both the added value of accounting for the across-link dynamics for traffic management and the added value of taking stochasticity in consideration

Ongoing work investigates formulations to approximate the full joint network distribution. For instance, we plan to combine the proposed approach with ideas from  the network decomposition method of \citep{FloOso14_DTA}. We are also investigating the use of aggregation-disaggregation techniques to approximate the full joint network distribution in a scalable and tractable way. Aggregation-disaggregation ideas  address the curse of dimensionality by providing an aggregate description of network states \cite{OsorioYamani_forthcoming,OsorioWang_submitted}.


\newpage

\bibliographystyle{trb}
\bibliography{biblio_network_latest}
\end{document}